\documentclass[fleqn,usenatbib]{mnras}

\usepackage{newtxtext,newtxmath}

\usepackage[T1]{fontenc}

\DeclareRobustCommand{\VAN}[3]{#2}
\let\VANthebibliography\thebibliography
\def\thebibliography{\DeclareRobustCommand{\VAN}[3]{##3}\VANthebibliography}

\usepackage{graphicx}	
\usepackage{amsmath}	

\newcommand{\angstrom}{\text{\normalfont\AA}}

\title[Mildly relativistic motion in PG~1351+640]{Mildly Relativistic Motion in the Radio Quiet Quasar PG~1351+640}

\author[Wang et al.]{
Ailing Wang,$^{1,2}$
Tao An,$^{1,2,3}$\thanks{E-mail: antao@shao.ac.cn}
Shaoguang Guo,$^{1,2,3}$
Luis C. Ho,$^{4,5}$
Willem A. Baan,$^{6,7}$
Robert Braun,$^{8}$ \newauthor
Sina Chen,$^{9}$
Xiaopeng Cheng,$^{10}$
Philippa Hartley,$^{8}$
Jun Yang,$^{11}$
Yingkang Zhang$^{1,3}$
\\
$^{1}$ Shanghai Astronomical Observatory, CAS, 80 Nandan Road, Shanghai 200030, China \\
$^{2}$ School of Astronomy and Space Sciences, University of Chinese Academy of Sciences, No. 19A Yuquan Road, Beijing 100049, China  \\
$^{3}$ Key Laboratory of Radio Astronomy and Technology, Chinese Academy of Sciences, A20 Datun Road, Chaoyang District, Beijing, 100101, P. R. China \\
$^{4}$ Kavli Institute for Astronomy and Astrophysics, Peking University, Beijing 100871, China \\
$^{5}$ Department of Astronomy, School of Physics, Peking University, Beijing 100871, China \\
$^{6}$ Xinjiang Astronomical Observatory, Chinese Academy of Sciences, 150 Science 1-Street, 830011 Urumqi, P.R. China \\
$^{7}$ Netherlands Institute for Radio Astronomy ASTRON, NL-7991 PD Dwingeloo, the Netherlands \\
$^{8}$ SKA Observatory, Jodrell Bank, Lower Withington, Macclesfield, SK11 9FT, UK \\
$^{9}$ Physics Department, Technion - Israel Institute of Technology, Haifa 3200003, Israel \\
$^{10}$ Korea Astronomy and Space Science Institute, 776 Daedeok-daero, Yuseong-gu, Daejeon 34055, Korea \\
$^{11}$ Department of Space, Earth and Environment, Chalmers University of Technology, Onsala Space Observatory, SE-439 92 Onsala, Sweden 
}

\date{Accepted XXX. Received YYY; in original form ZZZ}

\pubyear{2015}

\begin{document}
\label{firstpage}
\pagerange{\pageref{firstpage}--\pageref{lastpage}}
\maketitle

\begin{abstract}
Measuring the proper motion of the emission component in radio-quiet quasars (RQQs) could help to distinguish between the origins of the radio emission and to understand whether the jet production mechanism is the same in radio-loud quasars (RLQs) and RQQs. PG~1351+640 is one of the few RQQs suitable for proper motion studies: it has two compact components on milli-arcsecond scales, a flat-spectrum core and a steep-spectrum jet; both components are $\gtrsim$2 mJy at 5 GHz and are well suited for Very Long Baseline Array (VLBA) observations. We compare recent VLBA observations with that made seventeen years ago and find no significant change in the core-jet separation between 2005 and 2015 (a proper motion of 0.003 mas yr$^{-1}$). However,  the core-jet separation increased significantly between 2015 and 2022, inferring a jet proper motion velocity of 0.063 mas yr$^{-1}$, which corresponds to an apparent transverse velocity of $0.37\, c$. The result suggests that the jet of the RQQ PG 1351+640 is mildly relativistic and oriented at a relatively small viewing angle.
\end{abstract}

\begin{keywords}
galaxies: jets – galaxies: kinematics and dynamics – galaxies: ISM - quasars: individual (PG 1351+640).
\end{keywords}



\section{Introduction}

Quasars are commonly classified as radio-loud quasars (RLQs) or radio-quiet quasars (RQQs) based on their radio-to-optical flux density ratio $R$ (radio loudness parameter $R = S_{\rm 5GHz}$/$S_{4400\angstrom}$), with $R>10$ for RLQs and $R < 10$ for RQQs \citep{1989AJ.....98.1195K,1994AJ....108.1163K}. It is surprising that the optical properties of RLQs and RQQs are similar, despite significant differences in their observed radio properties (e.g., morphology, source size, luminosity, spectral index, etc.) \citep{1993MNRAS.263..425M,1989AJ.....98.1195K,2016ApJ...831..168K}. The radio emission of RLQs is thought to come from relativistic jets near the event horizon of the black hole \citep{2019ARA&A..57..467B}, but the origins of radio emission in RQQs are more  controversial. RQQ radio emission appears to be caused by a combination of star formation and AGN-related activities \citep[see][for a recent review]{2019NatAs...3..387P}.
Some studies have suggested that (at least some) RQQs are scaled-down versions of the more powerful RLQs \citep{1998MNRAS.299..165B,1996ApJ...471..106F,2005ApJ...618..108B,2005ApJ...621..123U} with the radio luminosity of  RQQs being three orders of magnitude lower than that of RLQs. 

Identifying relativistic jets in RQQs is crucial to understanding the dichotomy between RLQs and RQQs and whether the two groups have different types of jets. High-resolution radio imaging is a powerful tool to test whether RQQs have jets: if the radio emission of RQQs comes from relativistic jets, one would expect to observe a compact core with ultra-high brightness temperatures ($T_{\rm B}$) and/or a core-jet structure in the images. Very Long Baseline Interferometry (VLBI) observations of low-redshift RQQs with large radio flux densities (e.g. $>1$ mJy at 5 GHz) show that they do have compact core-jet structures or naked cores with high $T_{\rm B}$ \citep{1998MNRAS.299..165B,2005ApJ...621..123U,2022ApJ...936...73A,2023MNRAS.518...39W,2023PGpaper2}. This is observational evidence for the presence of relativistic jets in (some) RQQs, albeit with their relatively low radio power. Jets are more prevalent in RQQs with higher radio flux density or higher radio loudness. For example, relativistic jets have been detected in quasars with moderate $R$ ($10 < R < 250$, called radio-intermediate quasars, RIQs), which are thought to be relativistically boosted RQQs \citep{1993MNRAS.263..425M,1995A&A...298..375F,1996ApJ...471..106F}. 

Apparent superluminal motion is a typical characteristic of highly relativistic jets and is common in RLQs. Some RIQs also exhibit blazar-like variability, often accompanied by the formation of compact, short-lived jet knots (e.g. III Zw 2: \citealt{2000A&A...357L..45B,2023ApJ...944..187W}; Mrk 231: \citealt{2020ApJ...891...59R,2021MNRAS.504.3823W}). However, observations of jet proper motion in RQQs are still scarce, limiting the understanding of RQQ jets. A relativistic Doppler-boosted jet has only been observed in a few RQQ, such as PG~1407+263  \citep{2003ApJ...591L.103B}. Detecting relativistic jets with high or moderate relativistic velocities in more RQQs would help to investigate whether the central engine of RQQs is similar to that of RLQs.
In this Letter, we report the measurement of jet proper motion of PG~1351+640, a RQQ at $z = 0.088$\footnote{The following cosmological parameters are used throughout the paper: $\rm H_{0}=71 \, km \, s^{-1} \, Mpc^{-1}$, $\Omega_\Lambda = 0.73$ and $\Omega_{\rm m} = 0.27$. An angular size of 1 mas corresponds to a projected linear size of 1.626 parsec at the redshift of 0.088.} \citep{1988AJ.....95.1602S}. It is one of the ten RQQs detected in our Very Long Baseline Array (VLBA) observations \citep{2023MNRAS.518...39W,2023PGpaper2} and the only one in the sample showing a core and a compact jet component, giving us the rare opportunity to measure its jet motion from multi-epoch VLBI data.

\section{Methods, Observations and Data} \label{sec:method}

\begin{table*}
\caption{Image parameters. Column (1): observation date; Columns (2): observation code; Columns (3): observation frequency; Columns (4): observation bandwidth; Columns (5)-(7): major axis and minor axis of the restoring beam, and the position angle of the major axis, measured from north to east; Columns (8): peak flux density; Columns (9): root-mean-square (rms) noise of the image, measured in off-source regions.}
\begin{tabular}{lllrrllrc}
\hline \hline 
Obs. date  & Project Code & Freq. & Bandwidth & $B_{\rm maj}$   &$B_{\rm min}$  &$B_{\rm PA}$  &$S_{\rm peak}$  & $\sigma_{\rm rms}$   \\
(yyyy-mm-dd) &           &(GHz)  &(MHz)      & (mas)       & (mas)     &(\degr)   &(mJy beam$^{-1}$)    &(mJy beam$^{-1}$)   \\
(1)       & (2)          & (3)   & (4)       & (5)         & (6)       & (7)      & (8)        & (9)            \\
\hline 
2005-08-19 & BB203        & 5.0   & 32        & 1.72        & 1.01      & 143      & 3.83       & 0.026 \\
2015-08-04 & BA114B       & 5.0   & 256       & 3.75        & 1.95      & 6.21     & 2.51       & 0.029 \\
2022-01-22 & BW138A2      & 1.6   & 512       & 11.16       & 4.89      & 107      & 13.97      & 0.029 \\
2022-01-23 & BW138B2      & 4.7   & 256       & 4.22        & 1.68      & 110      & 2.65       & 0.028 \\
2022-01-23 & BW138B2      & 6.2   & 256       & 3.25        & 1.31      & 111      & 2.45       & 0.030 \\
2022-01-23 & BW138B2      & 7.6   & 256       & 2.62        & 1.04      & 110      & 2.56       & 0.032 \\ 
2022-03-05 & BC273G       & 5.0   & 256       & 4.16        & 1.83      & 30.6     & 2.13       & 0.032 \\                                                                                    \hline 

\end{tabular} \label{tab:obs}
\end{table*}

Table \ref{tab:obs} summarises the VLBA data used in this study. In addition to the new observations made on 22-23 January 2022 at frequencies of 1.6, 4.7, 6.2 and 7.6 GHz \citep{2023PGpaper2} and on 5 March 2022 (Chen et al. 2023, in prep.), we also include the published 4.9 GHz data observed on 4 August 2015 by \citet{2023MNRAS.518...39W} and archive data observed on 19 August 2005 from the National Radio Astronomy Observatory (NRAO) archive\footnote{NRAO data archive: \url{https://data.nrao.edu/portal/}}. 
Details of the observations and data processing for the 2015 and 2022 epochs are presented in \citet{2023MNRAS.518...39W,2023PGpaper2}.
For the 2005 data, we calibrated the raw data using the pipeline\footnote{The VLBI pipeline: \url{https://github.com/SHAO-SKA/vlbi-pipeline}} deployed at the China SKA Regional Centre \citep{2022SCPMA..6529501A}. After calibration, we imported the visibility data into the \textsc{Difmap} software package \citep{1997ASPC..125...77S} for mapping. 
We used the \textsc{modelfit} program in \textsc{Difmap} to obtain the VLBI component parameters, and fit the visibility data with two or three circular Gaussian models. The fitted parameters of the VLBI components are listed in Table \ref{tab:model}. 

\begin{table*}
    \centering
    \caption{Observational and model fitting results of the VLBI data.}
    \begin{tabular}{ccccccccc}
    \hline \hline 
    Obs. date     &Freq.  & Comp. & $S_{\rm int}$     & R                  & PA               &$\theta_{\rm FWHM}$ &log($T_{\rm B}$)   & Ref.    \\
    (yyyy-mm-dd) &(GHz)  &       & (mJy)             &(mas)               &($\degr$)         &(mas)               &log(K)             &         \\ 
    (1)          & (2)   & (3)   & (4)               & (5)                & (6)              & (7)                & (8)               & (9)     \\
    \hline 
    
    2005-08-19   &5.0    &SE (core) &0.62$\pm$0.03      &0                   &0               &$<$0.04           &$>$10.4                &1        \\
                 &       &NW (jet)  &4.68$\pm$0.23      &4.96$\pm$0.02       &$-$50.5$\pm$1.6 &0.57$\pm$0.03     & 8.88$\pm$0.07         &         \\
                 &       &J0 (jet)  &0.73$\pm$0.04      &2.68$\pm$0.01       &$-$47.6$\pm$1.6 &1.05$\pm$0.02     & 7.54$\pm$0.05         &         \\
    2015-08-04   &4.9    &SE (core) &1.98$\pm$0.20      &0                   &0               &0.28$\pm$0.02     & 9.13$\pm$0.19         & 2       \\
                 &       &NW (jet)  &3.22$\pm$0.16      &4.99$\pm$0.01       &$-$51.8$\pm$0.9 &1.19$\pm$0.02     & 7.76$\pm$0.06         &         \\ 
    2022-01-22   &1.6    &SE (core) &2.57$\pm$0.13      &0                   &0               &2.17$\pm$0.14     & 8.45$\pm$0.14         & 3       \\
                 &       &NW (jet)  &13.62$\pm$0.68     &3.89$\pm$0.02       &$-$48.8$\pm$0.8 &1.47$\pm$0.03     & 9.51$\pm$0.06         &         \\
    2022-01-23   &4.7    &SE (core) &2.43$\pm$0.12      &0                   &0               &0.25$\pm$0.05     & 9.38$\pm$0.43         & 3       \\
                 &       &NW (jet)  &3.02$\pm$0.15      &5.49$\pm$0.02       &$-$52.7$\pm$1.2 &0.90$\pm$0.04     & 8.36$\pm$0.11         &         \\
                 &       &J0 (jet)  &0.86$\pm$0.04      &2.85$\pm$0.07       &$-$53.5$\pm$1.4 &1.31$\pm$0.15     & 7.48$\pm$0.23         &         \\
    2022-01-23   &6.2    &SE (core) &2.46$\pm$0.12      &0                   &0               &$<$0.64           & $>$8.31               & 3       \\
                 &       &NW (jet)  &2.24$\pm$0.11      &5.34$\pm$0.01       &$-$53.3$\pm$0.7 &1.02$\pm$0.01     & 7.87$\pm$0.06         &         \\
                 &       &J0 (jet)  &0.38$\pm$0.03      &2.71$\pm$0.04       &$-$50.3$\pm$1.4 &0.90$\pm$0.07     & 7.21$\pm$0.17         &         \\       
    2022-01-23   &7.6    &SE (core) &2.85$\pm$0.14      &0                   &0               &0.35$\pm$0.03     & 8.73$\pm$0.19         & 3       \\
                 &       &NW (jet)  &1.27$\pm$0.06      &5.31$\pm$0.04       &$-$53.0$\pm$1.4 &0.92$\pm$0.07     & 7.53$\pm$0.16         &         \\
   2022-03-05   &5.0     &SE (core) &2.26$\pm$0.11      &0                   &0               &0.60$\pm$0.01     & 8.55$\pm$0.05         & 4       \\ 
                 &       &NW (jet)  &2.58$\pm$0.13      &5.32$\pm$0.01       &$-$52.7$\pm$1.6 &1.36$\pm$0.02     & 7.89$\pm$0.52         &         \\
                 &       &J0 (jet)  &0.24$\pm$0.03      &2.85$\pm$0.06       &$-$74.1$\pm$1.6 &$<$0.85           &$>$7.89                &         \\    
    \hline
   \end{tabular}\\
   Column (1) observation date;  (2) observing frequency;  (3) component label, the J0 component is located between the C and NW components;  (4) to (7) the model fitting parameters in sequence: the integrated flux density, radial separation with respect to the core, position angle (measured from north through east), component size (full width at half maximum of the fitted Gaussian component);  (8) brightness temperature. 
    (9) the reference of the dataset: Ref.~1 -- NRAO archive; Ref.~2 -- \citep{2023MNRAS.518...39W}; Ref.~3 -- \citep{2023PGpaper2}; Ref. 4 -- Chen et al. (2023, in prep.) 
   \label{tab:model}
\end{table*}

\section{Parsec-scale radio emission} \label{sec:result}

\begin{figure*}
\centering
\begin{tabular}{ccc}
\includegraphics[height=5cm]{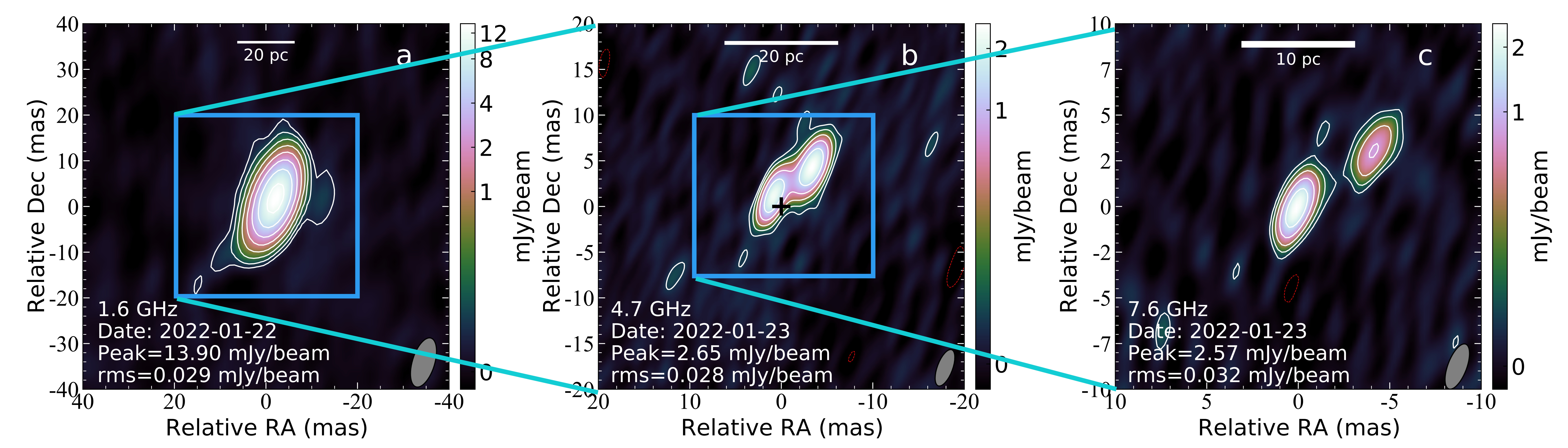}
\end{tabular}
  \caption{VLBA images of PG 1351+640 \citep{2023PGpaper2}. The image parameters are referred to Table~\ref{tab:obs} and are also labelled in each panel and the contours begin at 3 times rms noise and growth in a step of 2. The black cross indicates the position of the optical nucleus derived from the \textit{Gaia} data base (RA=13:53:15.8312, Dec.=63:45:45.6828). The uncertainty of the optical position is about 0.02 mas.  The radio position uncertainty mainly comes from the position error of the phase reference calibrator J1353+6324, which is 0.23 mas in both RA and Dec directions.  The annotations in the lower left corner indicate the observing frequency, observation date, peak flux density and root-mean-square (rms) noise of the image, respectively.
  }
  \label{fig:1351-map}
\end{figure*}

Figure \ref{fig:1351-map} shows the VLBA images in increasing order of frequency. Other images show a very similar morphology to that on 22/23 January 2022, so we do not repeat the presentation here. Figure \ref{fig:1351-map}-$a$ shows an unresolved source at 1.6 GHz. At 4.7 GHz and above (Fig. \ref{fig:1351-map}-$b$ and \ref{fig:1351-map}-$c$), the source is clearly resolved into two components. Column 8 of Table \ref{tab:model} lists the brightness temperatures estimated from the VLBI data. The lower limit to the  brightness temperature of the SE component is $T_{\rm B} = (3.8 \pm 0.3) \times 10^{8}$~K. The NW component has a slightly lower brightness temperature of  $T_{\rm B} = (1.8 \pm 0.1) \times 10^{8}$~K. The detection of compact, high brightness temperature radio components in VLBI images offers strong evidence that the parsec-scale radio emission is associated with the AGN. There are several sources that can produce high $T_{\rm B}$, such as jets or the corona. But the corona only can not interpret the resolved double-component structure. 

PG~1351+640 is highly variable in the radio band, with the total flux density at 5 and 15 GHz varying by more than  a factor of 4 from 1977 to 1986 \citep{1989ApJS...70..257B}. The VLBI components are also highly variable, with the 5-GHz flux density of SE increasing from $0.62\pm0.03$ mJy in August 2005 to $1.98\pm0.20$ mJy in August 2015, and further to $2.43 \pm 0.12$ mJy in January 2022 (Fig. \ref{fig:1351var}). 
The rapid and large variability is incompatible with the accretion disk wind scenario.
In addition, the two VLBI components have different spectral indices: NW has a steep spectrum between 1.6 and 7.6 GHz $\alpha_{1.6}^{7.6} = -1.45\pm0.12$ on epoch 23 January 2022 and is associated with optically-thin emission, while SE shows a flat spectrum $\alpha_{1.6}^{7.6} = -0.04\pm0.12$ and is associated with optically thick emission \citep{2023PGpaper2}. 
The resolved radio structure obtained from the VLBI data and the different spectral indices of the VLBI components support the notion that the parsec-scale radio emission of PG~1351+640 comes from a core-jet structure, that is, SE is the core (jet base) and NW is a jet knot. The position of the optical nucleus (marked by a cross in Fig.~\ref{fig:1351-map}-$b$), derived from the \textit{Gaia} data release 3 \citep{2022A&A...667A.148G}, is close to SE, reinforcing the identification of it as the radio core. 
Figure \ref{fig:1351-map}-$c$ shows the 7.6-GHz image, which is very similar to the 4.7-GHz image. The minor difference in morphology is that the steep-spectrum NW component becomes relatively weak at 7.6 GHz.
On the bridge between the SE and NW components there is a weak and steep-spectrum component J0. It was detected in both the 2005 and 2022 observations but did not show any significant proper motion.

Figure \ref{fig:1351SED} shows the plot of flux density versus observation frequency for PG~1351+640. In addition to the data listed in Table \ref{tab:model}, we also included the VLBA data observed in February 2000 \citep{2005ApJ...621..123U}. The spectral index of the whole source, derived from the epoch 2000 data, is $\alpha_{1.4}^{5.0} = -0.91 \pm 0.04$ (\citealt{2005ApJ...621..123U}), which is consistent with the spectral index obtained from our data  $\alpha_{1.6}^{7.6} = -0.94\pm0.02$ (epoch January 2022). 
For comparison, the spectral index from the VLA data is $\alpha^{8.5}_{5.0} = -0.64 $ \citep{2019MNRAS.482.5513L}. These results consistently show that steep spectrum emission is the dominant contribution to the radio flux density between 1.4 and 7.6 GHz on both kiloparsec and parsec scales. The radio core (solid data points) exhibits a flat spectrum with a spectral index of  $\alpha_{1.6}^{7.6} = -0.04 \pm 0.12$ (epoch 2022), clearly different from the steep spectrum of the whole source (open black data points). 
A comparison of the 5 GHz VLBA and VLA data shows that about 30 per cent of the flux density is in the compact core-jet structure in the nuclear region. Although the VLBA and VLA observations are not conducted simultaneously, they suggest that a significant fraction of the extended emission at scales $>$20 parsec is resolved at mas resolutions.

\section{Discussion and conclusion} \label{sec:disc}
\begin{figure}
\centering
\includegraphics[width=0.45\textwidth]{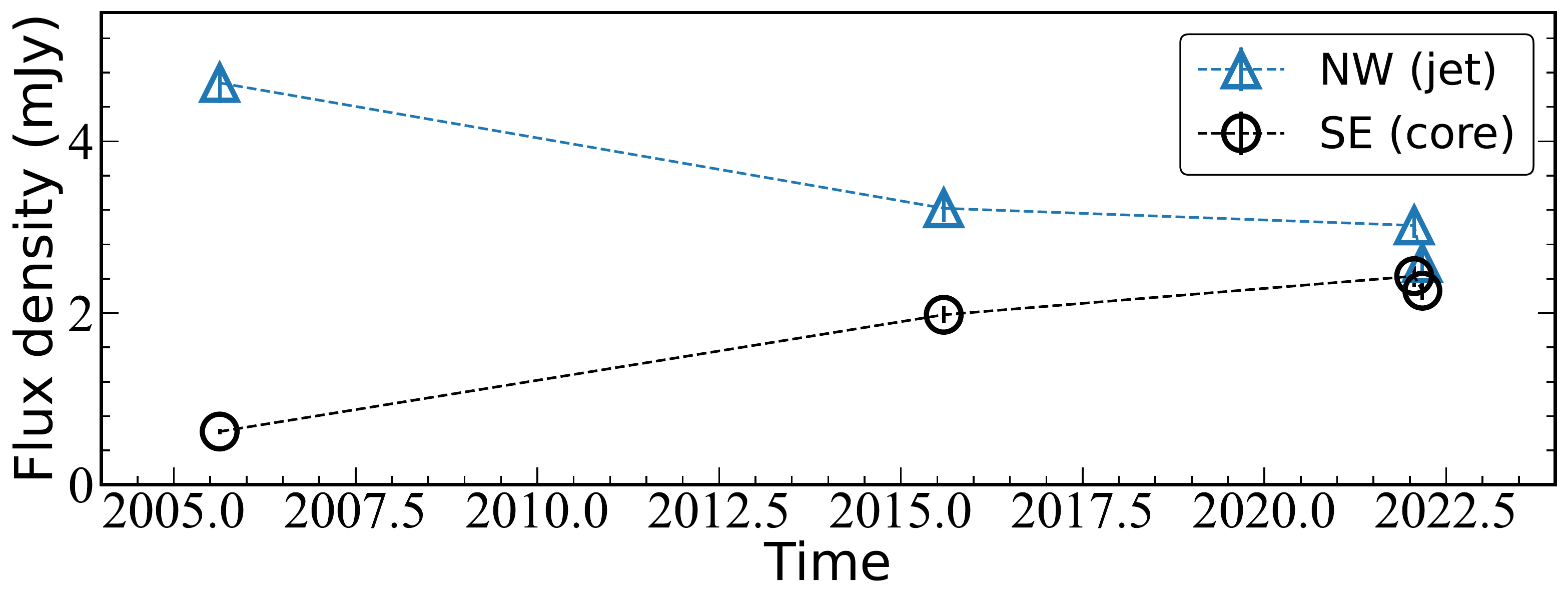}
  \caption{Change of flux densities of VLBI components of PG~1351+640 with time. The data are referred to Table \ref{tab:model}. }
  \label{fig:1351var}
\end{figure}

\begin{figure}
\centering
\includegraphics[width=0.45\textwidth]{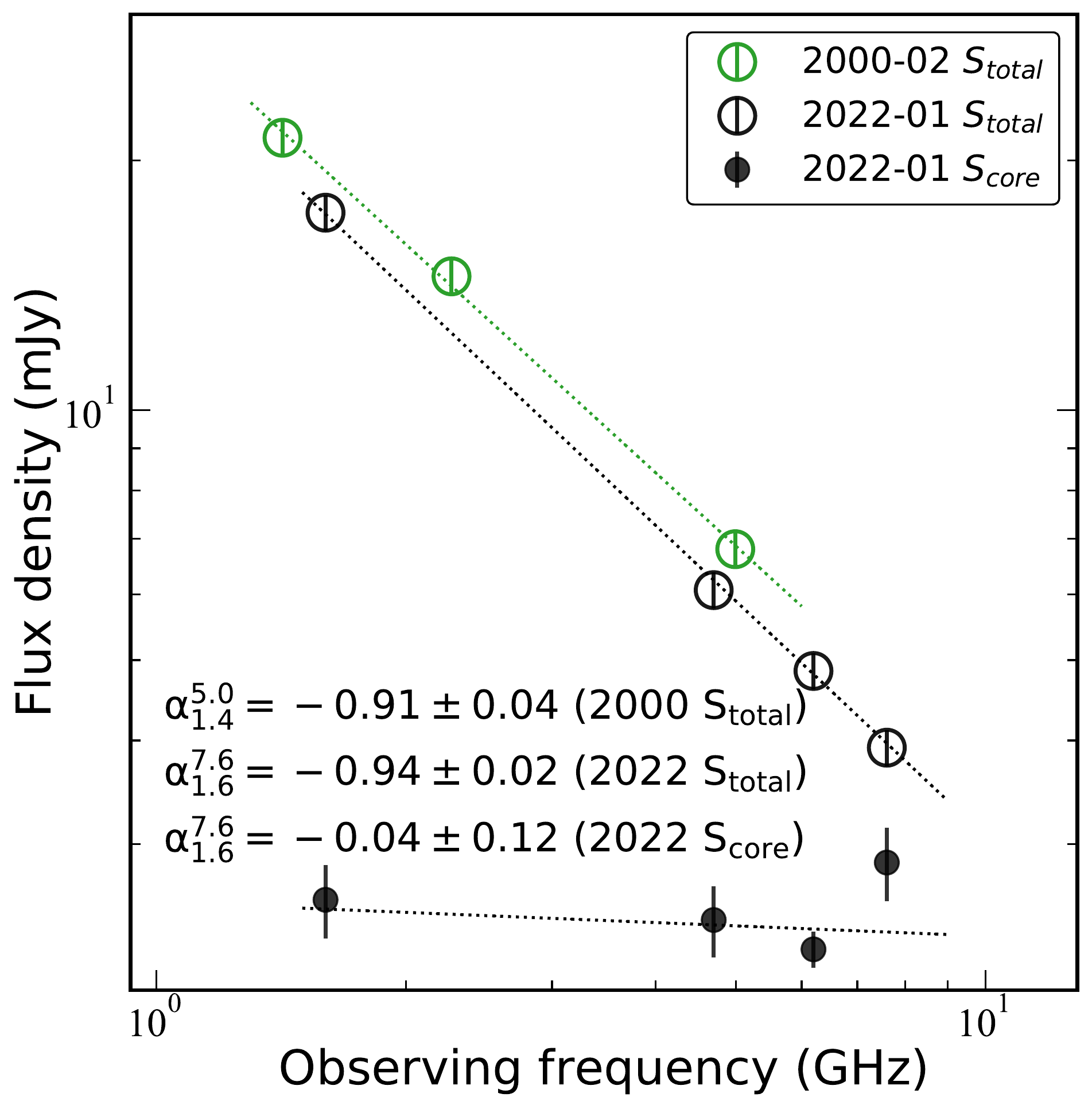}
  \caption{Flux densities of PG~1351+640 versus observing frequencies. The data points are obtained from VLBI observations, see details in Section \ref{sec:result}. The solid circular data point marks the flux density from the core component, while the open point represents the flux density of the entire VLBI source. The black data point of epoch 2022 is obtained from \citep{2023PGpaper2}, while the green point of epoch 2000 is taken from \citep{2005ApJ...621..123U}. }
  \label{fig:1351SED}
\end{figure}

Whether jets are prevalent in RQQs, whether RQQs have relativistic jets, and whether RQQ jets are produced by a mechanism similar to that of RLQ jets remain open questions.
Measuring or constraining the proper motion of emission components in RQQs can help distinguish the origin of radio emission from jet, wind or corona, and to study the jet properties. It should be noted that weak jets or misaligned jets with large viewing angles may not be detected, limited by the resolution and sensitivity of a given VLBA image.
In fact, in most VLBA images of RQQs only an unresolved core is detected \citep[e.g.,][]{2023MNRAS.518...39W}. PG~1351+640 is a rare case in which a compact jet knot is observed and clearly separated from the core.

\begin{figure}
\centering
\begin{tabular}{ccc}
\includegraphics[height=8cm]{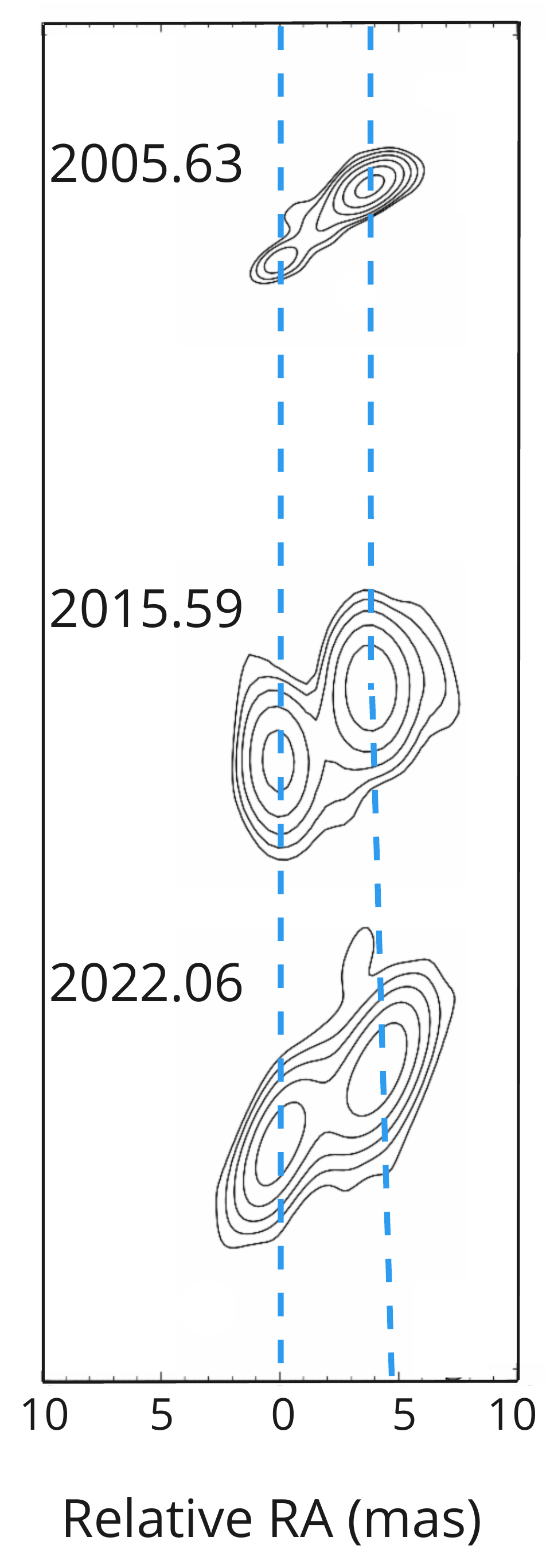} 
\includegraphics[width=0.3\textwidth]{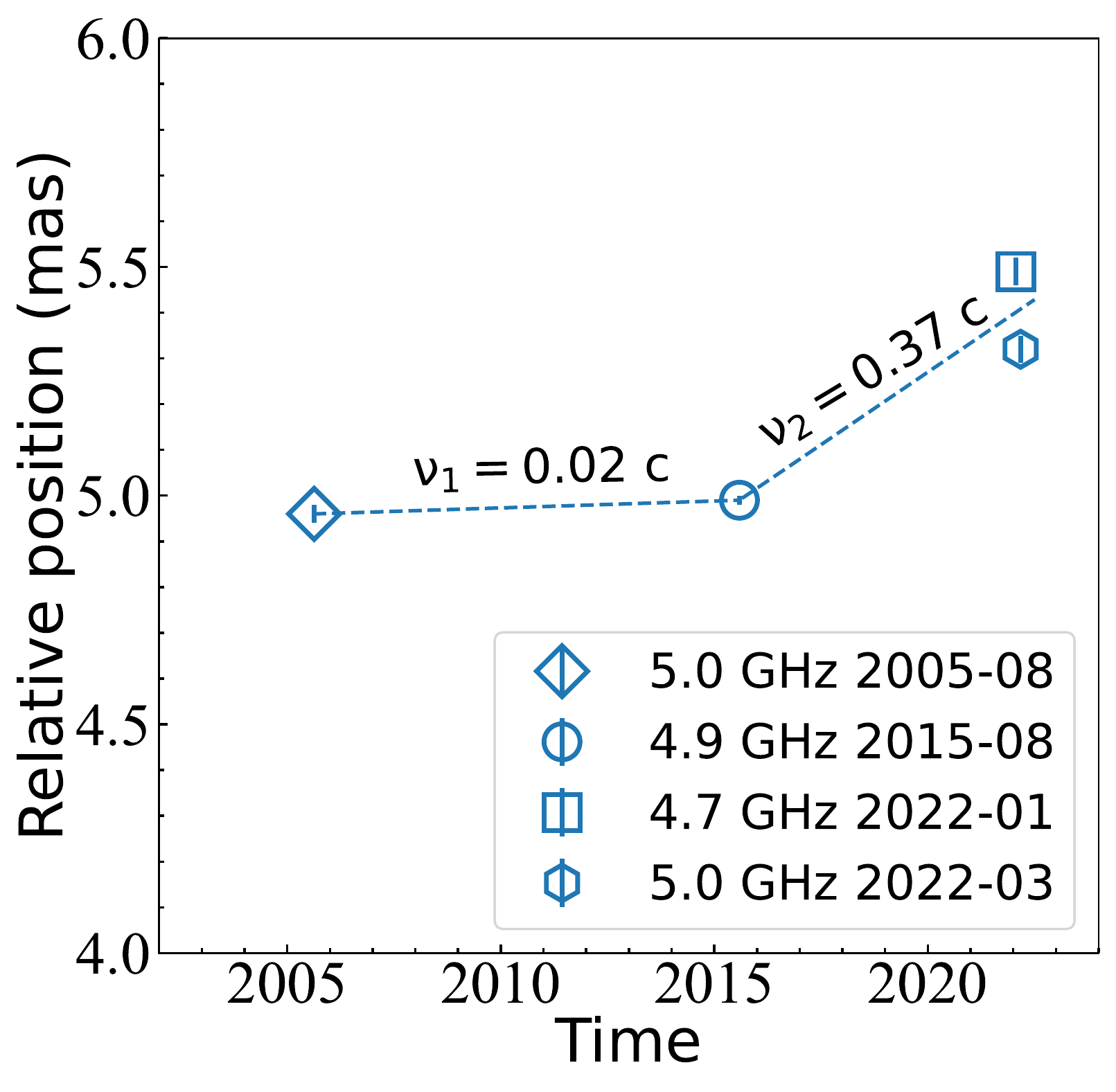}
\end{tabular}
  \caption{5 GHz VLBA images of PG~1351+640 with multiple-epoch at C-band revealing the proper motion (left) and the change of the core-jet distance with time (right). The jet velocity is $0.02\,c$ between 2005 and 2015, and $0.37\,c$ between 2015 and 2022. The data is related to Table \ref{tab:model}.
  }
  \label{fig:proper_motion}
\end{figure}

In many cases, the absolute position of the source is lost during the processing of the VLBI data. Therefore, we estimate the jet proper motion velocity by measuring the change of the distance of the jet with respect to the core over time. To avoid frequency-dependent opacity effects, we only use data from the same observed frequency (i.e. 5 GHz).
The proper motion obtained using all data points from 2005 to 2022 is 0.029$\pm$0.013 mas/yr, corresponding to $0.17 \pm 0.08$ c. The large uncertainty in the proper motion results from an apparent difference in the proper motion before and after 2015 (Fig. \ref{fig:proper_motion}): $\mu_1 = 0.003$ mas yr$^{-1}$ ($\beta_{\rm app,1} = 0.02$, 2005--2015, where $\beta_{\rm app}$ is the apparent transverse speed of the jet in units of light speed), $\mu_2  = 0.063$ mas yr$^{-1}$ ($\beta_{\rm app,2} = 0.37$, 2015--2022). 

A natural explanation for the non-detection of the counter jet in PG~1351+640 is the Doppler de-boosting effect. Assuming that the two-sided jets are identical and they are aligned with the line of sight at an angle of $\theta$. Assuming that the emission is isotropic in the jet rest frame, the brightness ratio between the jet and the counter jet can be calculated as $R_{\rm I} = I_{\rm j}/I_{\rm cj} = (\frac{1+\beta\cos\theta}{1-\beta\cos\theta})^{2+\alpha}$, where $\alpha$ is the spectral index of the jet component, $\beta$ is the jet speed in the unit of $c$. We adopt three times rms noise as an upper limit for $I_{\rm cj}$ and the peak flux density (see Table \ref{tab:obs}) for $I_{\rm j}$. Depending on the images obtained at different frequencies,  $R_{\rm I}$ ranges between 22 and 161. Another constraint on $\beta$ and $\theta$ comes from the measured apparent transverse velocity, i.e., $\beta_{\rm app} = \frac{\beta\sin\theta}{1 - \beta\cos\theta}$.
Substituting our observed $\beta_{\rm app,2}$ into the above calculation, we obtain $0.52<\beta<0.64$, $12\degr < \theta< 21\degr$. 
High-velocity blue-displaced absorption lines were observed in PG~1351+640 \citep{1994AJ....108.1178S}, and both the ultraviolet emission lines and continuum were found to be highly variable \citep{1985MNRAS.216..529T}. These observations indicate that the orientation of the broad line region of PG~1351+640 is close to the line of sight. In general, the estimate for the viewing angle of the jet axis of PG~1351+640 is consistent with the definition of a quasar in the AGN unification scheme \citep{1995PASP..107..803U}.

These results show that the jet of RQQ PG 1351+640 is mildly relativistic, assuming a relatively small viewing angle as expected for a quasar.
The jet proper motion velocity of PG 1351+640 is within the range of values obtained for the jet proper motion velocity of weak compact symmetric objects (CSOs) based on a larger sample \citep{2005ApJ...622..136G,2012ApJS..198....5A}.
The kinematic age of the parsec-scale of jet PG~1351+640 can be calculated between 90--180 years by using its proper motion speed and distance from the core.
However, it is important to note that this estimate is subject to several uncertainties. One major source of uncertainty is the measurement of the jet's proper motion speed. In the case of PG 1351+640, the proper motion calculation is based on only two epochs of data. To obtain a more accurate estimate of its kinematic age, additional VLBI observations would be needed. Another source of uncertainty is the assumption of a constant velocity for the hot spot. It is possible that the velocity of the hot spot has varied over time, which could affect the accuracy of the calculated kinematic age.
The de-projected size of the jet is $\sim63$ parsec. Placing PG~1351+640 in Fig. 7 of \citet{2012ApJ...760...77A}, we find that it is in an early evolutionary stage and the jet is still growing.

The mechanisms for the change in the jet proper motion are complex and may involve both intrinsic and extrinsic causes. One possible intrinsic mechanism is that shocks and instabilities in the jet flow lead to changes in the Lorentz factor of the propagating feature, which may affect the apparent motion of the jet \citep{2001ApJ...549L.183A}. This acceleration typically occurs within a projected distance of 10--20 parsec \citep{2015ApJ...798..134H}. The NW component of PG~1351+640 has a projection size of $\sim$13 parsec, which is in the region where the instability in the propagating shock plays a role. The other possible factors are extrinsic. If the jet axis is aligned to the line of sight with a small angle, any minor change in the viewing angle could cause notable changes in the jet proper motion. This phenomenon is more commonly seen in highly relativistic jets in blazars \citep{2015ApJ...798..134H}, but seems unlikely to be important for low-power jets in RQQs. Moreover, if the observed acceleration is due to a change in the jet direction, then a significant acceleration in the perpendicular jet direction should be observed as well, but is not seen in PG~1351+640.
Furthermore, the interaction of the jet with the external medium is actually responsible for the advance of the terminal hot spot. The observed change in the hot spot velocity may be a reflection of the change in the ambient medium. One possible scenario is that before 2015 the jet was blocked by a dense cloud that not only caused a deceleration of the NW hot spot but also increased its brightness (Fig. \ref{fig:1351var}); after 2015, this cloud was disrupted and the hot spot velocity returned to its original higher value. This hypothesis still needs to be tested with further observations.

In summary, PG~1351+640 is one of the few RQQs with direct proper motion measurements. Relativistic jets have been observed in several RQQs including PG~1351+640 \citep[e.g.][]{1998MNRAS.299..165B,2001ApJ...562L.133U,2003ApJ...591L.103B,2005ApJ...621..123U,2005ApJ...618..108B,2019MNRAS.485.3009H}, but with significantly different properties from the highly relativistic jets of RLQs. Both weak underlying jet flow and bright intermittent jet knots are detected in quasars with moderate radio loudness parameters and higher radio flux densities \citep{1995A&A...293..665F,1996ApJ...473L..13F,2000A&A...357L..45B,2020ApJ...891...59R,2021MNRAS.504.3823W,2023MNRAS.518...39W}, whose radio properties seem to between RQQs and RLQs. 
The nature and properties of the jets in RQQs, RIQs and RLQs appear to be related to the radio loudness  $R$. This means that the jet power and the degree of the relativistic beaming effect increase systematically from the RQQs to the RIQs to the RLQs as $R$ increases. Understanding the physical reasons for this evolutionary change will be the subject of further research.

\section*{Acknowledgements}
We are grateful to the anonymous referee for  constructive comments which help to improve the quality of the manuscript. AW and TA thank Kenneth Kellermann for the helpful discussion on the observational results and valuable comments on the manuscript.
This work is supported by the National SKA Program of China (grant number 2022SKA0120102). 
SG is supported by Youth Innovation Promotion Association CAS (2021258). 
LCH was supported by the National Science Foundation of China (11721303, 11991052, 12011540375, 12233001) and the China Manned Space Project (CMS-CSST-2021-A04, CMS-CSST-2021-A06).
SC acknowledges support by the Israel Science Foundation (grant no.1008/18) and a Center of Excellence of the Israel Science Foundation (grant no.2752/19).
The National Radio Astronomy Observatory is a facility of the National Science Foundation operated under cooperative agreement by Associated Universities, Inc. Scientific results from data presented in this publication are derived from the following VLBA project codes: BA114, BB203, BC273G, BW138.
The VLBI data processing made use of the computing resource of the China SKA Regional Centre. 

\section*{Data Availability}

The archival data used in this paper are available in the VLBA data archive (\url{https://archive.nrao.edu/archive/archiveproject.jsp}). The calibrated visibility data underlying this article can be requested from the corresponding author.



\bibliographystyle{mnras}
\bibliography{ref} 








\bsp	
\label{lastpage}
\end{document}